\documentclass[twocolumn,showpacs,prb,preprintnumbers,amsmath,amssymb]{revtex4}
\usepackage[latin1]{inputenc}
\usepackage{graphicx}
\usepackage{dcolumn}
\usepackage{bm}

\begin{document}

\title{Insulating spin liquid in the lightly doped two-dimensional Hubbard model}
\author{Hermann Freire}
\email{hermann@iccmp.org}
\author{Eberth Corrêa}
\author{Álvaro Ferraz}
 \affiliation{Centro Internacional de Física da Matéria Condensada, Universidade de Brasília, Brasília 04513, Brazil}
 \date{\today}

\begin{abstract}
We calculate the charge compressibility and uniform spin
susceptibility for the two-dimensional (2D) Hubbard model slightly
away from half-filling within a two-loop renormalization group
scheme. We find numerically that both those quantities flow to zero
as we increase the initial interaction strength from weak to
intermediate couplings. This result implies gap openings in both
charge and spin excitation spectra for the latter interaction
regime. When this occurs, the ground state of the lightly doped 2D
Hubbard model may be interpreted as an insulating spin liquid as
opposed to a Mott insulating state.
\end{abstract}


\maketitle

After two decades of intensive research on the high-Tc
superconductors, physicists are still puzzled by some of their very
unusual electronic properties \cite{Millis}. The prominent example
is given by the cuprates. At zero doping, despite the fact that
their highest occupied band is half-filled, they are charge
insulators, and display antiferromagnetic long-range order. For this
reason, they are said to be Mott insulators. As soon as one starts
doping those compounds with holes, the long-range magnetic order
becomes rapidly suppressed, and there are experimental evidences of
an emergence of a spin gap in their corresponding excitation spectra
\cite{Timusk}. A charge gap is also observed by ARPES experiments in
such lightly-doped systems \cite{Shen}. Moreover, at finite
temperatures, they turn themselves into poor conductors with
electronic properties differing considerably from the predictions of
Landau's Fermi liquid theory. This scenario configures the so-called
pseudogap regime. Although this phase continues to be not well
understood, it is widely acknowledged to play a fundamental role in
the underlying microscopic mechanism of such high-Tc
superconductors. Indeed, upon some further doping, those poor metals
become superconducting with $d$-wave symmetry up to relatively high
temperatures around the optimal doping level.

From the theoretical viewpoint, it is widely accepted that the
appropriate model for describing such systems is the two-dimensional
(2D) Hubbard model (HM), since it is known to have a Mott insulating
phase at half-filling, and is expected to become a $d$-wave
superconductor at larger doping \cite{Zanchi}. However, its
intermediate doping regime, which could provide some insight to
understand the physical nature of the pseudogap state, still remains
elusive to this date. In this Letter we intend to address this
question using renormalization group (RG) techniques in order to
infer about the ground state of such model for electron densities
slightly away from the half-filling limit.

\begin{figure}[b]
  \includegraphics[width=3.1in]{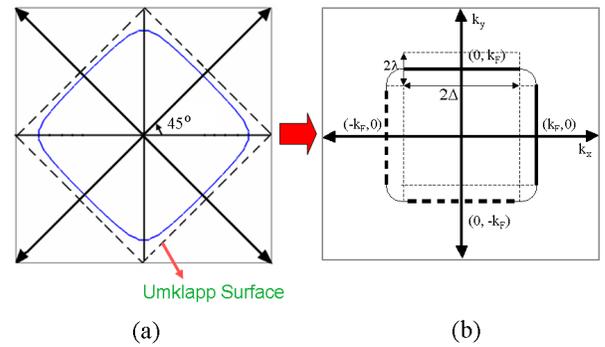}\\
  \caption{(Color online) (a) The half-filled Fermi surface (FS) of the 2D Hubbard model (dashed line), the lightly hole-doped FS (blue solid line), and
  (b) the latter FS after a rotation of the axes by $45^{o}$ degrees.}\label{fig1}
\end{figure}

Our considerations here will be based on a complete two-loop RG
calculation of the uniform charge (CS) and spin (SS)
susceptibilities of the 2D HM, taking into account simultaneously
both the renormalization of the couplings, and the self-energy
effects. (The CS is also called the charge compressibility of the
system.) To best of our knowledge, it is the first time that such a
full two-loop RG calculation is performed for the 2D lightly-doped
HM, since previous estimates of the uniform susceptibilities
followed random phase approximation (RPA) schemes.

In momentum space, the 2D Hubbard Hamiltonian on a square lattice is
given by

\begin{equation}
H=\sum_{\mathbf{k},\sigma}\xi_{\mathbf{k}}\psi^{\dagger}_{\mathbf{k}\sigma}\psi_{\mathbf{k}\sigma}+\left(\frac{U}{N_{sites}}\right)
\sum_{\mathbf{p},\mathbf{k},\mathbf{q}}\psi^{\dagger}_{\mathbf{p}+\mathbf{k}-\mathbf{q}\uparrow}\psi^{\dagger}_{\mathbf{q}\downarrow}
\psi_{\mathbf{k}\downarrow}\psi_{\mathbf{p}\uparrow},
\end{equation}

\noindent where the energy dispersion is simply
$\xi_{\mathbf{k}}=-2t\left[\cos(k_{x}a)+\cos(k_{y}a)\right]-\mu$,
and $\psi^{\dagger}_{\mathbf{k}\sigma}$ and
$\psi_{\mathbf{k}\sigma}$ are the usual creation and annihilation
operators of electrons with momentum $\mathbf{k}$ and spin
projection ${\sigma}=\uparrow,\downarrow$. Besides, $\mu$ stands for
the chemical potential, whereas $a$ is the square lattice spacing.
Another important parameter here is the width of the noninteracting
band, which is given by $W=8t$. This model describes a system with
many electrons interacting mutually via a local repulsive
interaction $U$, and with a total number $N_{sites}$ of lattice
sites. The electron band filling of the system is controlled by the
ratio $\mu/t$. When $\mu/t=0$ the system is exactly at half-filling.
As we start doping it with holes, $\mu/t$ takes slightly negative
values.

Our starting point is a 2D nearly flat Fermi surface (FS) with no
van Hove singularities (see Fig. 1). It correctly describes the 2D
HM slightly away from the half-filling case. A similar FS has
already been used by other groups to investigate the leading
instabilities within either a parquet or, equivalently, a one-loop
RG approach \cite{Yakovenko}. All those investigations find
diverging susceptibilities at finite energies (or finite
temperatures) with the dominant instability being always the spin
density wave (SDW). Their interpretation is that this implies a
spontaneous symmetry breaking in the system, and the onset of a
long-range ordered antiferromagnetic state.

In this Letter we argue that is not necessarily true since this
result may also be an indicative of the limitations of the one-loop
RG scheme. In low-dimensional systems, large quantum fluctuations
are expected to suppress long-range order. The more those effects
are taken into account, the more likely those long-range ordered
states are transformed into short-range magnetically ordered phases.
This also becomes clear as a result of our work. Taking into account
quantum fluctuation effects up to two-loop order in our RG scheme,
we are able to show that, for moderate coupling regimes, both charge
compressibility and uniform spin susceptibility are strongly
suppressed and flow unequivocally to zero. This behavior implies
that there are gaps for both charge and spin excitations, and no
trace of long-range symmetry breaking order in those cases. Such a
state with a fully gapped charge and spin spectra is usually
denominated an insulating spin liquid (ISL). The ISL is an example
of a short-range resonant valence bond state, which was first
proposed for a S=1/2 Heisenberg model \cite{Anderson}, and clearly
revealed in even-leg Hubbard ladders by both RG and bosonization
approaches \cite{Schulz,Balents,Dagotto}.

In order to implement a full RG calculation of the uniform
susceptibilities, it is essential to consider at least two-loop
order contributions. This is due to the fact that, at one-loop
level, there is not a single infrared (IR) divergent diagram in the
calculation of the so-called uniform response functions. In
contrast, there are several of those diagrams in two loops (the
nonparquet diagrams), and, as a result, one can reliably begin to
derive appropriate RG flow equations for those quantities at that
order of perturbation theory.

\begin{figure}[t]
  \includegraphics[width=2.4in]{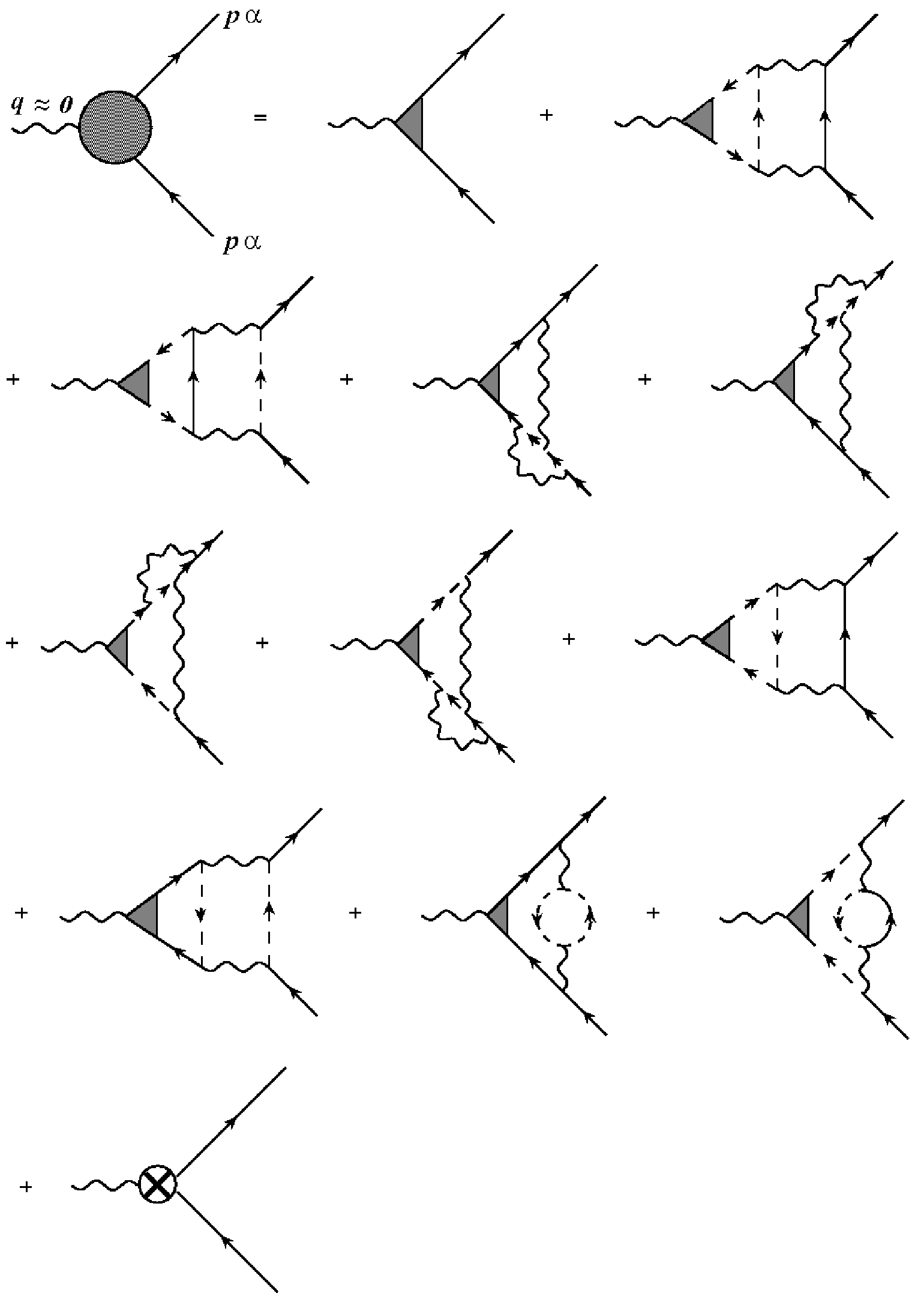}\\
  \caption{The uniform response function $\Gamma^{(2,1)}(\mathbf{p},\mathbf{q}\approx0)$ up to two-loop order. The triangular vertices stand for
  the renormalized response function, whereas the diagram with a cross represents the counterterm.}\label{fig2}
\end{figure}

To keep a closer contact with well-known works in one-dimensional
systems \cite{Solyom}, we divide the FS into four different regions
(two sets of solid and dashed line patches). Here we restrict the
momenta at the FS to the flat parts only. The interaction processes
connecting parallel patches of the FS are always logarithmically IR
divergent due to quantum fluctuations. In contrast, those connecting
perpendicular patches always remain finite, and do not contribute to
the RG flow equations in our approach. For convenience, we restrict
ourselves to one-electron states labeled by the momenta
$p_{\parallel}=k_{x}$ and $p_{\perp}=k_{y}$ associated with one of
the two sets of perpendicular patches. The momenta parallel to the
FS are restricted to the interval $-\Delta\leqslant p_{
\parallel}\leqslant\Delta$, with $2\Delta$ being
essentially the size of the flat patches. The energy dispersion of
the single-particle states is given by
$\varepsilon_{a}\left(\mathbf{p}\right)=v_{F}\left(\left|p_{\perp}\right|-k_{F}\right)$,
and depends only on the momenta perpendicular to the FS. The label
$a=\pm$ refers to the flat sectors at $p_{\perp}=\pm k_{F}$,
respectively. In addition, we take
$k_{F}-\lambda\leqslant\left|p_{\perp}\right|\leqslant
k_{F}+\lambda$, where $\lambda$ is a fixed ultraviolet (UV)
microscopic momentum cut-off.

We now write down the Lagrangian of the 2D HM as

\begin{eqnarray}
L=&&\sum_{\mathbf{p},\sigma,a=\pm}\psi_{(a)\sigma}^{\dagger}\left(\mathbf{p}\right)
\left[i\partial_{t}-v_{F}\left(\left|p_{\perp}\right|-k_{F}\right)\right]\psi_{(a)\sigma}
\left(\mathbf{p}\right)\nonumber \\
&&-\frac{1}{V}\sum_{\mathbf{p,q,k}}
\sum_{\alpha,\beta,\delta,\gamma}\left[g_{2}\delta_{\alpha\delta}\delta_{\beta\gamma}-g_{1}\delta_{\alpha\gamma}\delta_{\beta\delta}\right]\nonumber
\\&&\times\psi_{\left(+\right)\delta}^{\dagger}\left(\mathbf{p+q-k}\right)\psi_{\left(-\right)\gamma}^{\dagger}\left(\mathbf{k}\right)
\psi_{\left(-\right)\beta}\left(\mathbf{q}\right)\psi_{\left(+\right)\alpha}\left(\mathbf{p}\right),\nonumber\\
\label{lagrangian}
\end{eqnarray}

\noindent where $\psi^{\dagger}_{(\pm)}$ and $\psi_{(\pm)}$ are now
fermionic fields associated to electrons located at the $\pm$
patches. The summation over momenta must be appropriately understood
as $\sum_{\mathbf{p}}=V/(2\pi)^{2}\int d^{2}\mathbf{p}$ in the
thermodynamic limit. We linearized the energy dispersion about the
lightly-doped FS, and the interaction term was parametrized in a
manifestly SU(2) invariant form. Here we follow the well-known
g-ology notation, with $g_{1}$ and $g_{2}$ standing for
backscattering and forward scattering couplings, respectively. Since
this should represent the 2D HM, these couplings must be initially
defined as $g_{1}=g_{2}=(V/4N_{sites})U$. In addition, we do not
include Umklapp processes since we are not at half-filling, and our
FS does not intersect the so-called Umklapp surface  at any point
(see Fig. 1(a)). Consequently, our result is different from another
evidence of ISL behavior reported in the literature in the context
of the $t-t'$ 2D HM \cite{Furukawa}. In their case, the FS
intersects the Umklapp surface even away from half-filling, and, for
this reason, Umklapp processes become an essential ingredient for
the correct description of the system at the lightly-doped regime.
Moreover, their estimates of the uniform susceptibilities follow the
already mentioned RPA-like scheme.

\begin{figure}[b]
  \includegraphics[width=1.9in]{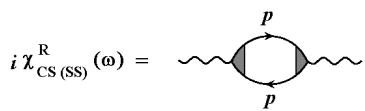}\\
  \caption{Feynman diagram associated with the renormalized charge compressibility and uniform spin susceptibility.}\label{fig3}
\end{figure}

Since the HM is a microscopic model, all terms in the Lagrangian are
defined at a scale of a few lattice spacings in real space (i.e at
the UV cutoff scale in momentum space). These parameters are often
inaccessible to every day experiments, since the latter probe only
the low-energy and the long-wavelength dynamics of a given system.
In RG theory, these unobserved quantities are known as the bare
parameters. In fact, if one attempts to construct naive perturbative
calculations with such parameters, one will obtain IR divergent
Feynman diagrams in the computation of several quantities such as,
e.g. the backscattering and the forward scattering four-point
vertices, and the single-particle Green's function. These
divergences mean that the perturbation theory setup is not
appropriately formulated. To solve this problem, one can redefine
the perturbation scheme in a such a way as to circumvent these
infinite results in the calculation of observable quantities. This
is the strategy of the so-called renormalized perturbation theory.
Thus, one rewrites all unobserved bare parameters in terms of the
associated observable renormalized (or physical) quantities. The
difference between them is given by a counterterm, whose fundamental
role is to cancel, by construction, the related IR divergences in
all orders of perturbation theory. If this program is successfully
accomplished, then the theory is said to be properly renormalized.

As we explained in our previous paper \cite{Hermann}, when one deals
with this FS problem, the counterterms needed to renormalize the
theory turn out to be continuous functions of the three momenta
parallel to the FS, rather than being simply infinite constants. In
addition, we computed the RG flow equations for the coupling
functions and the quasiparticle weight up to two-loop order, and
showed that they were in fact coupled integro-differential
equations. We solved those equations self-consistently and, as a
result, we found out for an intermediate coupling regime a possibly
new physical regime, which was characterized by a strongly
suppressed quasiparticle weight. Here we continue to explore this
interacting regime, and our present calculation shows that the
resulting quantum state may in fact be interpreted as an ISL.

To obtain the uniform susceptibilities of this system, we must first
calculate the linear response due to an infinitesimal external field
perturbation that couples to both charge and spin number operators.
We do this by adding to the Lagrangian the new term

\begin{equation}
-h_{external}\sum_{\mathbf{p},a=\pm}\mathcal{T}_{B}^{\alpha\alpha}(\mathbf{p})\psi_{(a)\alpha}^{B
\dagger}\left(\mathbf{p}\right)\psi^{B}_{(a)\alpha}\left(\mathbf{p}\right),
\end{equation}

\noindent where $B$ stands for the bare quantities. This will
generate an additional vertex (the one-particle irreducible function
$\Gamma^{(2,1)}(\mathbf{p},\mathbf{q}\approx0)$), which will in turn
be afflicted by new IR divergences (see the nonparquet diagrams in
Fig. 2). As a result, we must rewrite the bare quantity
$\mathcal{T}_{B}^{\alpha\alpha}$ in terms of its renormalized
counterpart (henceforth called $\mathcal{T}_{R}^{\alpha\alpha}$),
and an appropriate counterterm $\Delta
\mathcal{T}_{R}^{\alpha\alpha}$, i.e.

\begin{equation}
\mathcal{T}_{B}^{\alpha\alpha}(p_{\parallel})=Z^{-1}(p_{\parallel})\left[\mathcal{T}_{R}^{\alpha\alpha}(p_{\parallel})+\Delta
\mathcal{T}_{R}^{\alpha\alpha}(p_{\parallel})\right].
\end{equation}

\noindent The quasiparticle weight $Z$ factor comes from the
renormalization of the fermionic fields, which must be also taken
into account in order to include the feedback of the self-energy
effects into the RG flow equations. The Z function is calculated
explicitly in Ref. \cite{Hermann}. As was mentioned before, $\Delta
\mathcal{T}_{R}^{\alpha\alpha}$ must cancel exactly the divergences
generated by the nonparquet diagrams. However, there are still
several ways of choosing that counterterm. To solve this ambiguity,
we must make a prescription establishing precisely that the
$\mathcal{T}_{R}^{\alpha\alpha}$ is the experimentally observable
response, i.e.,
$\Gamma^{(2,1)}(p_{\parallel},p_{0}=\omega,p_{\perp}=k_{F};\textbf{q}\approx0)=-i\mathcal{T}_{R}^{\alpha\alpha}(p_{\parallel},\omega)$,
where $\omega$ is the RG energy scale parameter that denotes the
proximity of the renormalized theory to the FS. In this way, to flow
towards the FS we let $\omega\rightarrow0$.

We are now ready to define the two different types of uniform
response functions, which arise from a symmetrization with respect
to the spin projection, namely

\begin{eqnarray}
\mathcal{T}_{R,CS}(p_{\parallel},\omega)&=&\mathcal{T}_{R}^{\uparrow\uparrow}(p_{\parallel},\omega)+\mathcal{T}_{R}^{\downarrow\downarrow}(p_{\parallel},\omega),\\
\mathcal{T}_{R,SS}(p_{\parallel},\omega)&=&\mathcal{T}_{R}^{\uparrow\uparrow}(p_{\parallel},\omega)-\mathcal{T}_{R}^{\downarrow\downarrow}(p_{\parallel},\omega),
\end{eqnarray}

\noindent where $\mathcal{T}_{R,CS}$ and $\mathcal{T}_{R,SS}$ are
the response functions associated with the charge compressibility
and spin susceptibility, respectively. In order to compute the RG
flow equations for these response functions, one needs to recall
that the bare parameters are independent of the renormalization
scale $\omega$. Thus, using the RG condition $\omega
d\mathcal{T}_{B}^{\alpha\alpha}/d\omega=0$, we obtain

\begin{eqnarray}
\omega\frac{d}{d\omega}\mathcal{T}_{R,CS}(p_{\parallel})&=&-\omega\frac{d}{d\omega}\Delta\mathcal{T}_{R,CS}(p_{\parallel})
+\gamma(p_{\parallel})\mathcal{T}_{R,CS}(p_{\parallel}),\nonumber \\
\\
\omega\frac{d}{d\omega}\mathcal{T}_{R,SS}(p_{\parallel})&=&-\omega\frac{d}{d\omega}\Delta\mathcal{T}_{R,SS}(p_{\parallel})
+\gamma(p_{\parallel})\mathcal{T}_{R,SS}(p_{\parallel}), \nonumber
\\
\end{eqnarray}

\noindent where the anomalous dimension is given by
$\gamma(p_{\parallel})=\omega d \ln Z(p_{\parallel})/d \omega$.
Despite their apparent simplicity, it is impossible to solve these
RG equations only by analytical means. To find their solutions, we
have again to resort to numerics. Here we use the fourth-order
Runge-Kutta numerical method. We discretize the FS continuum
replacing the interval $-\Delta \leqslant p_{\parallel} \leqslant
\Delta$ by a discrete set of 33 points. We use $\omega =
\Omega\exp(-l)$, where $\Omega = 2v_{F}\lambda$ with $l$ being our
RG step. We also choose $\Omega=v_{F}\Delta < W$. In view of our
choice of points for the FS, we are only allowed to go up to
$l\approx2.8$ in the RG flow to avoid the distance to the FS being
smaller than the shortest distance between points in our discrete
set.

\begin{figure}[t]
  \includegraphics[width=3.2in]{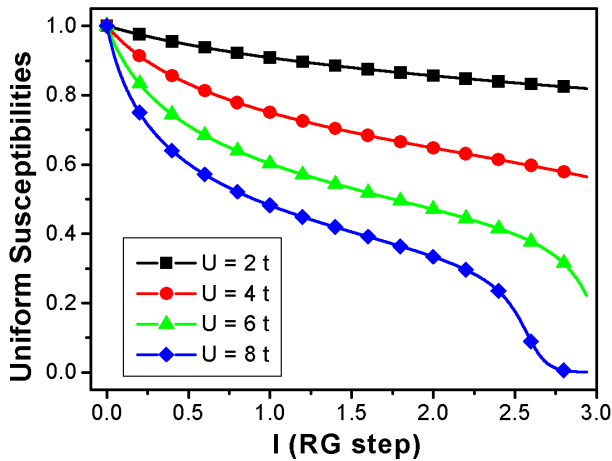}\\
  \caption{(Color online) RG flow of both charge compressibility and uniform spin susceptibility as we increase the initial coupling strength given by
  $U=(g/\pi v_{F})t$.}\label{fig4}
\end{figure}

Once the response functions are obtained, we can calculate the flow
of the charge compressibility and the uniform spin susceptibility of
the system. Using again our diagrammatic convention, it follows from
Fig. 3 that they are given by

\begin{equation}
\chi^{R}_{CS(SS)}(\omega)=\frac{1}{4\pi^{2}v_{F}}\int_{-\Delta}^{\Delta}dp_{\parallel}\left[\mathcal{T}_{R,CS(SS)}(p_{\parallel},\omega)\right]^2.\\
\end{equation}

\noindent To evaluate these quantities we follow the same numerical
procedure described above. Our results are displayed in Fig. 4. We
note in this plot that both charge and spin susceptibilities flow at
the same rate as we approach the FS even though their corresponding
response functions have different flow equations. In addition, for
initial couplings in which the quasiparticle weight flows to zero,
the uniform susceptibilities become strongly suppressed in the
low-energy limit. While the former asserts that there are no
fermionic quasiparticle excitations present in the system, the
latter is related to the complete absence of low-lying bosonic
charge and spin collective excitations. Since this resulting state
has only gapful excitations, it cannot be related to any broken
symmetry state and, as a consequence, should possess only
short-range ordering. This quantum state has, therefore, strong
similarities with that predicted long ago by Anderson
\cite{Anderson}, which is commonly referred to as an ISL. In our
present work, it becomes evident that such a state is produced by
disordering effects induced by strong quantum fluctuations, and
these are approximately taken into account by our two-loop RG
scheme. Finally, we call attention to the fact that an insulating
behavior in the lightly doped 2D HM was also reported recently in
the literature \cite{Phillips}. Our present result is clearly in
agreement with their results.

In summary, we examined the flow of both charge compressibility and
uniform spin susceptibility in the lightly doped 2D HM as a function
of the initial interaction strength within a two-loop RG approach.
For moderate interaction regimes, both quantities flow to zero as we
approach the initial FS of the system. This is a strong indicative
that there are gaps in both charge and spin excitation spectra of
the lightly doped 2D HM. Hence the quantum state associated with
that regime may be viewed as an ISL as opposed to a Mott insulator.
This result may be of relevance for the cuprate high-Tc
superconductors in view of the fact that the 2D HM in the
intermediate coupling regime is widely believed to be appropriate to
describe such compounds in all doping ranges.

This work was supported by the Conselho Nacional de Desenvolvimento
Científico e Tecnológico (CNPq).

\end{document}